\documentclass[aps,prl,twocolumn,superscriptaddress,showpacs,preprintnumbers,amsmath,amssymb]{revtex4-1}
\usepackage{graphicx} 
\usepackage{dcolumn}  

\usepackage[colorlinks, urlcolor=black]{hyperref}
\usepackage[hyphenbreaks]{breakurl}

\graphicspath{{ps}}

\begin{document}



\title{ \quad\\[1.0cm] Measurement of time-dependent $CP$ violation\\ in $B^0 \to K^0_S \pi^0 \pi^0$ decays}

\noaffiliation
\affiliation{University of the Basque Country UPV/EHU, 48080 Bilbao}
\affiliation{Beihang University, Beijing 100191}
\affiliation{Brookhaven National Laboratory, Upton, New York 11973}
\affiliation{Budker Institute of Nuclear Physics SB RAS, Novosibirsk 630090}
\affiliation{Faculty of Mathematics and Physics, Charles University, 121 16 Prague}
\affiliation{Chonnam National University, Kwangju 660-701}
\affiliation{University of Cincinnati, Cincinnati, Ohio 45221}
\affiliation{Deutsches Elektronen--Synchrotron, 22607 Hamburg}
\affiliation{Duke University, Durham, North Carolina 27708}
\affiliation{Key Laboratory of Nuclear Physics and Ion-beam Application (MOE) and Institute of Modern Physics, Fudan University, Shanghai 200443}
\affiliation{Justus-Liebig-Universit\"at Gie\ss{}en, 35392 Gie\ss{}en}
\affiliation{Gifu University, Gifu 501-1193}
\affiliation{SOKENDAI (The Graduate University for Advanced Studies), Hayama 240-0193}
\affiliation{Gyeongsang National University, Chinju 660-701}
\affiliation{Hanyang University, Seoul 133-791}
\affiliation{University of Hawaii, Honolulu, Hawaii 96822}
\affiliation{High Energy Accelerator Research Organization (KEK), Tsukuba 305-0801}
\affiliation{J-PARC Branch, KEK Theory Center, High Energy Accelerator Research Organization (KEK), Tsukuba 305-0801}
\affiliation{Forschungszentrum J\"{u}lich, 52425 J\"{u}lich}
\affiliation{IKERBASQUE, Basque Foundation for Science, 48013 Bilbao}
\affiliation{Indian Institute of Science Education and Research Mohali, SAS Nagar, 140306}
\affiliation{Indian Institute of Technology Bhubaneswar, Satya Nagar 751007}
\affiliation{Indian Institute of Technology Guwahati, Assam 781039}
\affiliation{Indian Institute of Technology Hyderabad, Telangana 502285}
\affiliation{Indian Institute of Technology Madras, Chennai 600036}
\affiliation{Indiana University, Bloomington, Indiana 47408}
\affiliation{Institute of High Energy Physics, Chinese Academy of Sciences, Beijing 100049}
\affiliation{Institute of High Energy Physics, Vienna 1050}
\affiliation{INFN - Sezione di Napoli, 80126 Napoli}
\affiliation{INFN - Sezione di Torino, 10125 Torino}
\affiliation{Advanced Science Research Center, Japan Atomic Energy Agency, Naka 319-1195}
\affiliation{J. Stefan Institute, 1000 Ljubljana}
\affiliation{Institut f\"ur Experimentelle Teilchenphysik, Karlsruher Institut f\"ur Technologie, 76131 Karlsruhe}
\affiliation{King Abdulaziz City for Science and Technology, Riyadh 11442}
\affiliation{Department of Physics, Faculty of Science, King Abdulaziz University, Jeddah 21589}
\affiliation{Korea University, Seoul 136-713}
\affiliation{Kyungpook National University, Daegu 702-701}
\affiliation{LAL, Univ. Paris-Sud, CNRS/IN2P3, Universit\'{e} Paris-Saclay, Orsay}
\affiliation{\'Ecole Polytechnique F\'ed\'erale de Lausanne (EPFL), Lausanne 1015}
\affiliation{P.N. Lebedev Physical Institute of the Russian Academy of Sciences, Moscow 119991}
\affiliation{Faculty of Mathematics and Physics, University of Ljubljana, 1000 Ljubljana}
\affiliation{Ludwig Maximilians University, 80539 Munich}
\affiliation{Luther College, Decorah, Iowa 52101}
\affiliation{University of Maribor, 2000 Maribor}
\affiliation{Max-Planck-Institut f\"ur Physik, 80805 M\"unchen}
\affiliation{School of Physics, University of Melbourne, Victoria 3010}
\affiliation{University of Mississippi, University, Mississippi 38677}
\affiliation{University of Miyazaki, Miyazaki 889-2192}
\affiliation{Moscow Physical Engineering Institute, Moscow 115409}
\affiliation{Moscow Institute of Physics and Technology, Moscow Region 141700}
\affiliation{Graduate School of Science, Nagoya University, Nagoya 464-8602}
\affiliation{Universit\`{a} di Napoli Federico II, 80055 Napoli}
\affiliation{Nara Women's University, Nara 630-8506}
\affiliation{National Central University, Chung-li 32054}
\affiliation{National United University, Miao Li 36003}
\affiliation{Department of Physics, National Taiwan University, Taipei 10617}
\affiliation{H. Niewodniczanski Institute of Nuclear Physics, Krakow 31-342}
\affiliation{Nippon Dental University, Niigata 951-8580}
\affiliation{Niigata University, Niigata 950-2181}
\affiliation{Novosibirsk State University, Novosibirsk 630090}
\affiliation{Osaka City University, Osaka 558-8585}
\affiliation{Pacific Northwest National Laboratory, Richland, Washington 99352}
\affiliation{Panjab University, Chandigarh 160014}
\affiliation{Peking University, Beijing 100871}
\affiliation{University of Pittsburgh, Pittsburgh, Pennsylvania 15260}
\affiliation{Punjab Agricultural University, Ludhiana 141004}
\affiliation{Theoretical Research Division, Nishina Center, RIKEN, Saitama 351-0198}
\affiliation{University of Science and Technology of China, Hefei 230026}
\affiliation{Showa Pharmaceutical University, Tokyo 194-8543}
\affiliation{Soongsil University, Seoul 156-743}
\affiliation{University of South Carolina, Columbia, South Carolina 29208}
\affiliation{Sungkyunkwan University, Suwon 440-746}
\affiliation{School of Physics, University of Sydney, New South Wales 2006}
\affiliation{Department of Physics, Faculty of Science, University of Tabuk, Tabuk 71451}
\affiliation{Tata Institute of Fundamental Research, Mumbai 400005}
\affiliation{Department of Physics, Technische Universit\"at M\"unchen, 85748 Garching}
\affiliation{Toho University, Funabashi 274-8510}
\affiliation{Department of Physics, Tohoku University, Sendai 980-8578}
\affiliation{Earthquake Research Institute, University of Tokyo, Tokyo 113-0032}
\affiliation{Department of Physics, University of Tokyo, Tokyo 113-0033}
\affiliation{Tokyo Institute of Technology, Tokyo 152-8550}
\affiliation{Tokyo Metropolitan University, Tokyo 192-0397}
\affiliation{Virginia Polytechnic Institute and State University, Blacksburg, Virginia 24061}
\affiliation{Wayne State University, Detroit, Michigan 48202}
\affiliation{Yamagata University, Yamagata 990-8560}
\affiliation{Yonsei University, Seoul 120-749}
  \author{Y.~Yusa}\affiliation{Niigata University, Niigata 950-2181} 
  \author{H.~Aihara}\affiliation{Department of Physics, University of Tokyo, Tokyo 113-0033} 
  \author{S.~Al~Said}\affiliation{Department of Physics, Faculty of Science, University of Tabuk, Tabuk 71451}\affiliation{Department of Physics, Faculty of Science, King Abdulaziz University, Jeddah 21589} 
  \author{D.~M.~Asner}\affiliation{Brookhaven National Laboratory, Upton, New York 11973} 
  \author{H.~Atmacan}\affiliation{University of South Carolina, Columbia, South Carolina 29208} 
  \author{V.~Aulchenko}\affiliation{Budker Institute of Nuclear Physics SB RAS, Novosibirsk 630090}\affiliation{Novosibirsk State University, Novosibirsk 630090} 
  \author{T.~Aushev}\affiliation{Moscow Institute of Physics and Technology, Moscow Region 141700} 
  \author{R.~Ayad}\affiliation{Department of Physics, Faculty of Science, University of Tabuk, Tabuk 71451} 
  \author{V.~Babu}\affiliation{Tata Institute of Fundamental Research, Mumbai 400005} 
  \author{I.~Badhrees}\affiliation{Department of Physics, Faculty of Science, University of Tabuk, Tabuk 71451}\affiliation{King Abdulaziz City for Science and Technology, Riyadh 11442} 
% ...
\collaboration{The Belle Collaboration}

\begin{abstract}
We report a measurement of time-dependent $CP$ violation in $B^0 \to K^0_S \pi^0 \pi^0$ decays using a data sample of $772 \times 10^6$ $B\bar{B}$ pairs collected by the Belle experiment running at the $\Upsilon (4S)$ resonance at the KEKB $e^+ e^-$ collider. This decay proceeds mainly via a $b\to sd\bar{d}$ ``penguin'' amplitude. The results are
$\sin 2\phi^{\rm eff}_1 = 0.92^{+0.27}_{-0.31}~$ (stat.) $\pm 0.11$ (syst.) and $\mathcal{A} = 0.28 \pm 0.21$ (stat.) $\pm 0.04$ (syst.), which are the most precise measurements of $CP$ violation in this decay mode to date.
The value for the $CP$-violating parameter $\sin 2\phi^{\rm eff}_1$ is consistent with that obtained using decay modes proceeding via a $b\to c\bar{c}s$ ``tree'' amplitude.
\end{abstract}


\maketitle

\tighten

{\renewcommand{\thefootnote}{\fnsymbol{footnote}}}
\setcounter{footnote}{0}

In the Standard Model (SM), $CP$ violation in the quark sector is induced by a complex phase in the Cabibbo-Kobayashi-Maskawa (CKM) quark mixing matrix \cite{KM}.　At $\Upsilon (4S) \to B\bar{B}$ transitions, for neutral $B$ meson decays into a $CP$ eigenstate produced, the decay rate has a time dependence \cite{Sanda, CPVrev}
\begin{eqnarray}
\label{eqn_dt_signal}
\mathcal{P}(\Delta t,q) & = & \frac{e^{-|\Delta t|/\tau_{B^0}}}{4\tau_{B^0}}\times \\ 
 & & \biggl( 1 + q\bigl[ \mathcal{S}\sin(\Delta m^{}_d\Delta t) 
+ \mathcal{A}\cos(\Delta m^{}_d\Delta t)  \bigr] \biggr),\nonumber
\end{eqnarray}
where $\mathcal{S}$ and $\mathcal{A}$ are $CP$-violating parameters; $q=1$ for $\bar{B}^0$ decays and $-1$ for $B^0$ decays; $\Delta t$ is the difference in decay times of the $B^0$ and $\bar{B}^0$ mesons; $\Delta m^{}_d$ is the mass difference between the two mass eigenstates of the $B^0$-$\bar{B}^0$ system; and $\tau^{}_{B^0}$ is the $B^0$ lifetime.
As the $B^0 \to K^0_S \pi^0 \pi^0$ decays proceeds mainly via a $b\to sd\bar{d}$ ``penguin'' amplitude, and the final state is $CP$ even \cite{plb596_163}, the SM expectation is $\mathcal{S}\approx -\sin 2\phi_1$ and $\mathcal{A}\approx 0$, where $\phi_1 =$ arg$[(-V_{cd}V_{cb}^*)/(V_{td}V_{tb}^*)]$. Deviations from these expectations could indicate new physics. The value of $\sin 2\phi^{}_1$ is well-measured using decays proceeding via a $b\to c\bar{c}s$ tree amplitude, and thus comparing our measurement of $\sin 2\phi^{\rm eff}_1$ to the $b\to c\bar{c}s$ value \cite{phi1_belle, phi1_babar} provides a test of the SM \cite{phi1_NP}. We note that there is a $b\to u\bar{u}s$ tree amplitude that also contributes to $B^0 \to K^0_S \pi^0 \pi^0$ decays and can shift $\phi^{\rm eff}_1$ from $\phi^{}_1$; however, this amplitude is doubly Cabibbo-suppressed, and thus the resulting shift is very small \cite{shift_btou}. Previously, the BaBar experiment studied this decay and measured $\sin 2\phi^{\rm eff}_1= -0.72 \pm 0.71 \pm 0.08$ \cite{babar2007}; here we present the first such measurement from the Belle experiment using a data sample 3.4 times larger than that of BaBar.

%
The Belle detector is a large-solid-angle magnetic
spectrometer that consists of a silicon vertex detector (SVD),
a 50-layer central drift chamber (CDC), an array of
aerogel threshold Cherenkov counters (ACC),  
a barrel-like arrangement of time-of-flight
scintillation counters (TOF), and an electromagnetic calorimeter
comprised of CsI(Tl) crystals (ECL) located inside 
a super-conducting solenoid coil that provides a 1.5~T
magnetic field.  An iron flux-return located outside 
the coil is instrumented to detect $K_L^0$ mesons and to identify
muons (KLM).  The detector
is described in detail elsewhere~\cite{Belle}.
Two inner detector configurations were used. A 2.0 cm radius beampipe
and a 3-layer silicon vertex detector was used for the first sample
of $152 \times 10^6 B\bar{B}$ pairs, while a 1.5 cm radius beampipe, a 4-layer
silicon detector and a small-cell inner drift chamber were used to record  
the remaining $620 \times 10^6 B\bar{B}$ pairs \cite{svd2}.  

Due to the asymmetric energies of the $e^+$ and $e^-$ beams, the $\Upsilon(4S)$ is produced with a Lorentz boost of $\beta\gamma=0.425$ nearly parallel to the $+z$ axis, which is defined as the direction opposite the $e^+$ beam. Since the $B^0 \bar{B}^0$ pair is almost at rest in the $\Upsilon(4S)$ center-of-mass (CM) frame, the decay time difference $\Delta t$ can be determined from the separation along $z$ of the $B^0$ and $\bar{B}^0$ decay vertices: $\Delta t\approx (z_{CP} - z_{\rm tag})/(\beta\gamma c)$, where $z_{CP}$ and $z_{\rm tag}$ are $z$-coordinates of the decay positions of the $B^0$ decaying to the $CP$ eigenstate and the other (tag-side), respectively. To reconstruct the decay vertices without the presence of primary charged tracks, we extrapolate the reconstructed $K^0_S$ momentum back to the region of the interaction point (IP) and use the IP profile in the transverse plane (perpendicular to the $z$ axis) as a constraint. This method was used in a previous Belle analysis of $B^0 \to K^0_S \pi^0$ decays \cite{pi0k0-belle} and is described in detail in Ref. \cite{ksksks-belle}. Compared to $B^0 \to K^0_S \pi^0$ decays, the $K^0_S$ in three-body $B^0\to K^0_S \pi^0 \pi^0$ decays has lower momentum and thus tends to decay closer to the IP; this results in about a 20\% larger yield of $K^0_S$ decays to $\pi^+ \pi^-$ inside the SVD volume with a correspondingly higher vertex reconstruction efficiency and greater precision in the $B$ decay vertex position as discussed in Ref. \cite{plb596_163}.

In the determination of the event selection, Monte Carlo simulated events (MC) are used. For the signal, 1 million events for each of non-resonant, $K^*(892)^0 \pi^0$ and $f_0 K^0_S$ all of which decay into $K^0_S \pi^0\pi^0$ final state, are generated using the EVTGEN \cite{Evtgen} event generator package. These resonant states are also $CP$-eigenstates induced by same diagram as the non-resonant decay. Using these MC samples, all of the states are confirmed to be reconstructed and not to be affected by the selections. For the background, a large number of $B\bar{B}$ and $q\bar{q}$ processes are simulated. Interactions of the particles in the Belle detector are reproduced using GEANT3 \cite{GEANT} with detector configuration information in each time period of the experiment.

Candidate $K^0_S$ decays are selected using multivariate analysis based on a neural network technique \cite{neurobayse, nisks}. The input variables to select displaced vertices are as follows: the distance between two daughter pion tracks in the $z$ direction; the flight distance in the $x$-$y$ plane; the angle between the momentum of the $\pi^+ \pi^-$ system and the $K^0_S$ candidate's vertex position vector with respect to the IP; and the shortest distance between the IP and daughter tracks of the $K^0_S$ candidate. In addition, we use the momenta of the $K^0_S$ and $\pi$, the angle between the $K^0_S$ and $\pi$, and hit information of daughters in the SVD and CDC. In this analysis we require that candidates satisfy the selection 0.480 GeV/$c^2 < M_{\pi^+ \pi^-} <$ 0.516 GeV/$c^2$, where $M_{\pi^+ \pi^-}$ is the reconstructed invariant mass of the charged pions. This range corresponds to  approximately 3$\sigma$ in the resolution of the mass.

Candidate $\pi^0 \to \gamma \gamma$ decays are reconstructed using photon candidates identified from ECL hits. We require that $M_{\gamma \gamma}$ satisfy 0.115 GeV/$c^2 < M_{\gamma \gamma} <$ 0.152 GeV/$c^2$, which corresponds to approximately 3$\sigma$ in resolution. To improve the $\pi^0$ momentum resolution, we perform a mass-constrained fit to the two photons, assuming they originate from the IP.

In the case of multiple $B^0$ candidates in an event, we select the candidate that combines the $\pi^0$ of the smallest mass-constrained fit $\chi^2$ value with the $K^0_S$ of the largest value of the neural network output variable.  

To identify the decay $B^0 \to K^0_S \pi^0 \pi^0$, we define two variables: the beam-constrained mass ${M_{\rm bc} \equiv \sqrt{(E_{\rm beam}/c^2)^2 - |\vec{p}_B^{\rm ~CM}/c|^{2}}}$, and the energy difference $\Delta E \equiv E_{\rm beam} - E^{\rm CM}_{B}$, where $\vec{p}^{\rm ~CM}_B$ and $E^{\rm CM}_{B}$ are the $B$ momentum and energy, respectively, in the $e^+ e^-$ CM frame. The quantity $E_{\rm beam}$ is the beam energy in the CM frame. The variables $M_{\rm bc}$ and $\Delta E$ for signal events peak at the $B^0$ mass and at zero, respectively, but have tails to lower values due to lost energy in the $\pi^0$ reconstruction.

To reject background $B\bar{B}$ decays resulting in the $K^0_S \pi^0 \pi^0$ final state, we define veto regions for the reconstructed invariant masses $M_{K^0_S \pi^0}$ and $M_{\pi^0 \pi^0}$. Decays $B^0 \to D^0 X$ and $B^0 \to K^0_S \pi^0$ are rejected by vetoing the regions 1.77 GeV/$c^2 < M_{K^0_S \pi^0}  <$ 1.94 GeV/$c^2$ and $M_{K^0_S \pi^0} > 4.8$ GeV/$c^2$, respectively, for both $\pi^0$ candidates individually combined with the $K^0_S$ candidate. The veto region for $B^0 \to (c\bar{c}) K^0_S$ is 2.8 GeV/$c^2 < M_{\pi^0\pi^0} <$ 3.6 GeV/$c^2$, where $(c\bar{c})$ is dominated by the charmonium mesons. Many of two-body decays of the $B^0$ into neutral meson and $K^0_S$ are $CP$ eigenstates. Among such decay modes, $B^0 \to \eta' K^0_S$ becomes background if photons are not detected with the decays of $\eta' \to \eta \pi^+ \pi^-$, $\eta \to 2\gamma$ and $K^0_S \to \pi^0 \pi^0$ so that $M_{\pi^0\pi^0} < 0.6$ GeV/$c^2$ is vetoed. In addition to those invariant masses of intermediate states, the absolute value of cosine of the angle between the photons and the $\pi^0$ boost direction of laboratory in the $\pi^0$ rest frame is required to be less than 0.9 to reject $B \to X_s \gamma$ decays, where $X_s$ denotes hadronic state governed by a radiative penguin decay.

\begin{figure}[htb]
\includegraphics[width=0.35\textwidth]{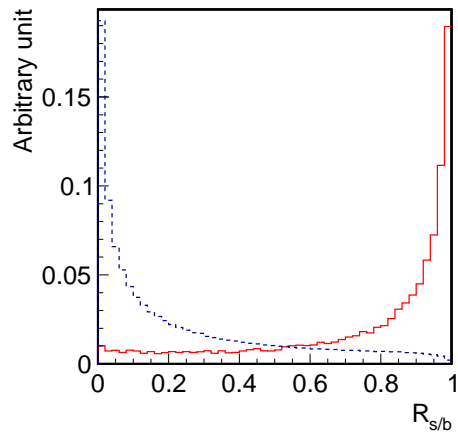}
\caption{Distribution of $\mathcal{R}_{\rm s/b}$, an event-shape based likelihood ratio, for signal and $q\bar{q}$ MC illustrated by solid and broken lines, respectively.}
\label{fig_signal_lhr}
\end{figure}

To suppress $e^+e^- \to q\bar{q}$ continuum background events, a likelihood ratio $\mathcal{R}_{\rm s/b}$ is calculated using modified Fox-Wolfram moments \cite{SFW, KSFW} and the cosine of the angle between the beam direction and $B^0$ flight direction in the CM frame, $\cos \theta_B$. Figure \ref{fig_signal_lhr} shows the $\mathcal{R}_{\rm s/b}$ distribution of the signal and $q\bar{q}$ MC. We impose a loose requirement $\mathcal{R}_{\rm s/b} > 0.50$, which rejects 84\% of continuum background while retaining 90\% of signal decays. We subsequently include a probability density function (PDF) for $\mathcal{R}_{\rm s/b}$ when fitting for the signal yield.

The vertex of the tag-side $B$ is reconstructed from all charged tracks in the event, except for the $K^0_S$ daughters, using a vertex reconstruction algorithm described in Ref. \cite{vertexres}. To determine the $B^0$ flavor $q$, a multi-dimensional likelihood-based method for inclusive properties of particles not associated with the signal $B^0$ candidate is used \cite{TaggingNIM}. The quality of the flavor tagging result is expressed by $r$, where $r = 0$ corresponds to no flavor discrimination, and $r = 1$ corresponds to unambiguous flavor assignment. Candidates with $r \le 0.10$ are not considered further for $CP$ volation measurement. The wrong tag fractions for six $r$ intervals, $w_l (l = 1$-6), and their differences between $B^0$ and $\bar{B}^0$ decays, $\Delta w_l$, are determined from large control samples of self-tagging $B^0 \to D^{*-} \ell^+ \nu, B^0 \to D^{(*)-} h^+ (h=\pi, \rho)$ decays. The total effective tagging efficiency defined as $\displaystyle \Sigma (f_l \times (1-2w_l)^2)$ is determined to be ($29.8 \pm 0.4$)\%, where $f_l$ is the fraction of the events in the $l$-th interval. 

\begin{figure}[htb]
\includegraphics[width=0.50\textwidth]{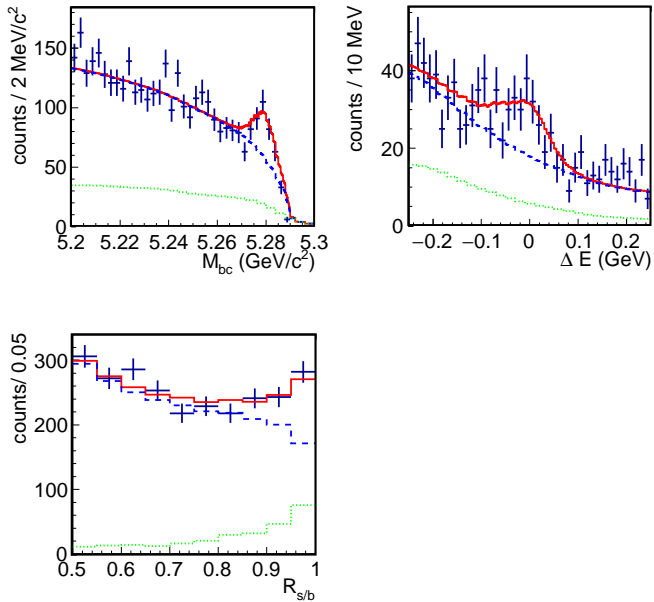}
\caption{$M_{\rm bc}$, $\Delta E$ and $\mathcal{R}_{\rm s/b}$ distributions (points with uncertainties) using signal-enhanced selections $M_{\rm bc} > 5.27$ GeV$/c^2$, $-0.15$ GeV $< \Delta E < 0.10$ GeV, and $\mathcal{R}_{\rm s/b} >0.9$ except for the variable displayed. The fit result is illustrated by the solid curve, while the total and $B\bar{B}$ backgrounds are shown by broken and dotted curves, respectively.}
\label{fig_sigext}
\end{figure}

After applying all selection criteria, the signal yield is extracted from a three-dimensional unbinned maximum likelihood fit to $M_{\rm bc}$, $\Delta E$, and $\mathcal{R}_{\rm s/b}$. For signal and $B\bar{B}$ background, the PDFs are modeled as binned histograms determined from MC simulation. A two-dimensional PDF is used for $M_{\rm bc}$ and $\Delta E$, taking into account the correlation between these variables. The $q\bar{q}$ background PDF for $M_{\rm bc}$ is modeled by an ARGUS function \cite{ARGUS}, and that for $\Delta E$ is modeled by second-order polynomial function. A binned histogram from the MC is used for the $q\bar{q}$ background PDF of $\mathcal{R}_{\rm s/b}$. From the 43225 events in the regions of  $M_{\rm bc} > 5.2$ GeV$/c^2$, $-0.25$ GeV $< \Delta E < 0.25$ GeV, and $\mathcal{R}_{\rm s/b} >0.5$, the yields of signal, $q\bar{q}$ and $B\bar{B}$ are found to be $335 \pm 37$, $38599 \pm 262$ and $4290 \pm 190$, respectively. Figure \ref{fig_sigext} shows the data distribution in the signal-enhanced region $M_{\rm bc} > 5.27$ GeV$/c^2$, $-0.15$ GeV $< \Delta E < 0.10$ GeV, and $\mathcal{R}_{\rm s/b} >0.9$, together with the fit projections, where the selection requirement on the plotted quantity is released.

\begin{figure}[htb]
\includegraphics[width=0.4\textwidth]{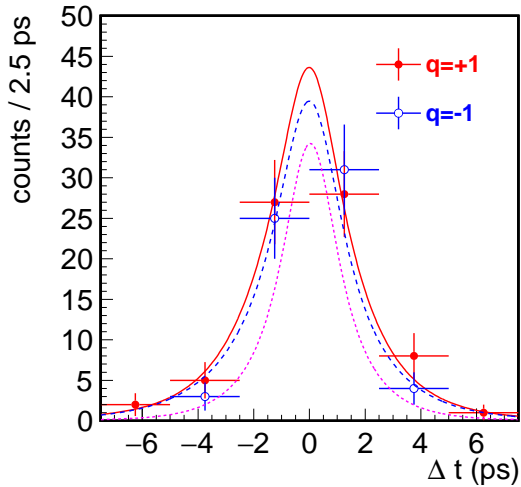}
\caption{$\Delta t$ distribution shown by data points with uncertainties, and fit result with curves: filled circles with error bars along with a solid-line fit curves correspond to $q=+1$, while open circles with error bars along with a dashed-line fit curves correspond to $q=-1$. The background contribution is illustrated by dotted-line. Events with good flavor tagging quality ($r>0.5$) are shown.}
\label{fig_delt}
\end{figure}

To measure the $CP$ violation parameters, an unbinned maximum likelihood fit is performed for the $\Delta t$ distribution using $q$ from the flavor tagging procedure and the signal fraction evaluated from the signal extraction fit. The PDF for the signal is set to take the form of Eq. \ref{eqn_dt_signal_det} which is obtained by modifying Eq. \ref{eqn_dt_signal} for wrong tagging and vertex resolution:
\begin{eqnarray}
  \displaystyle \mathcal{P}(\Delta t, q) =   \hspace*{7cm} \nonumber\\
   \frac{e^{-|\Delta t|/\tau_{B^0}}}{4\tau_{B^0}} \Big( 1 - q\Delta w
    + (1 - 2w) q \big[ \mathcal{S} \sin(\Delta m_d \Delta t)\hspace*{1cm} \nonumber \\
      + \mathcal{A} \cos(\Delta m_d \Delta t) \big] \Big) \otimes R(\Delta t),\hspace*{0.5cm}  
  \label{eqn_dt_signal_det}
\end{eqnarray}
where $R(\Delta t)$ is a convolved resolution function consisting of three components: the detector resolution for $z_{CP}$ and $z_{\rm tag}$ vertices; the shift of $z_{\rm tag}$ due to secondary tracks; and the kinematic approximation used in calculating $\Delta t$ from the vertex positions. These are determined using a large $CP$-conserving sample of semi-leptonic and hadronic $B$ decays.  For the background, which includes both $q\bar{q}$ and $B\bar{B}$, the PDF is modeled as a combination of two Gaussian functions and a delta function, as determined from the sideband regions 5.20 GeV$/c^2 < M_{\rm bc} < 5.26$ GeV$/c^2$, $-1.00$ GeV $< \Delta E < -0.40$ GeV and 0.20 GeV $< \Delta E < 0.50$ GeV. $\tau_B$ and $\Delta m_d$ are fixed to world average values \cite{PDG}. For the resolution function $R(\Delta t)$, a broad Gaussian function is included to account for a small outlier component. The number of events within the three-dimensional region of $M_{\rm bc} > 5.27$ GeV$/c^2$, $-0.15$ GeV $< \Delta E < 0.10$ GeV and $\mathcal{R}_{\rm s/b} >0.5$ with vertices and flavor information is 964, and the purity is 11.4\%. From fitting these events we obtain $\mathcal{S} = -0.92 ^{+0.31}_{-0.27}$ and $\mathcal{A} = 0.28 \pm 0.21$, where the errors are statistical only.　Figure \ref{fig_delt} shows the $\Delta t$ distribution of each flavor together with the background.

The systematic uncertainties are summarized in Table \ref{tab_syst}. Systematic uncertainties originating from vertexing opposite the $CP$ side, flavor tagging, and fixed physics parameters, and tag-side interference \cite{TSI} are estimated from studying the large statistic data sample of the ${B^0 \to (c\bar{c}) K^0}$ analysis \cite{phi1_belle}. Uncertainty from vertex reconstruction using $K^0_S$ including the resolution function is estimated using a large control sample of $B^0 \to J/\psi K^0_S$ decays. Fit bias is estimated by fitting a large number of signal MC samples and evaluating the resulting deviation compared to the input. For the PDF shape, the uncertainty is estimated using a smeared distribution. For parameters fixed in the fit, such as the signal fraction and background $\Delta t$ PDF, the uncertainties are estimated by shifting these parameters by their errors and refitting; the resulting changes in $\mathcal{S}$ and $\mathcal{A}$ are taken as the systematic uncertainties. Including the systematic uncertainty, we determine that $\sin 2 \phi^{\rm eff}_1 = 0.92~ ^{+0.27}_{-0.31} \pm 0.11$ and $\mathcal{A} = 0.28 \pm 0.21 \pm 0.04$, where first and second errors are statistic and systematic, respectively.
\begin{table}[htb]
\centering
\caption{Systematic uncertainties}
\label{tab_syst}
  \begin{tabular}
    {@{\hspace{0.5cm}}l@{\hspace{0.5cm}}|@{\hspace{0.5cm}}c@{\hspace{0.5cm}}|@{\hspace{0.5cm}}c@{\hspace{0.5cm}}}
    \hline \hline
                      & $\Delta \sin 2 \phi_1^{\rm eff}$ & $\Delta\mathcal{A}$\\
    \hline
Vertexing             & $\pm 0.02$         & $\pm 0.01$\\
Flavor tagging        & $\pm 0.004$        & $\pm 0.003$\\
Resolution function   & $\pm 0.06$   & $^{+0.004}_{-0.003}$\\
Physics parameters    & $\pm 0.002$        & $<0.001$\\
Fit bias              & $\pm 0.03$         & $\pm 0.02$\\
Background fraction   & $\pm 0.02$         & $\pm 0.02$\\
Background $\Delta t$ & $\pm 0.08$  & $\pm 0.02$\\
Tag-side interference & $\pm 0.001$        & $\pm 0.008$\\
\hline
Total                 & $\pm 0.11$ & $\pm0.04$\\
    \hline \hline
  \end{tabular}
\end{table}

In summary, we measure $CP$ violation parameters in the decay $B^0 \to K^0_S \pi^0 \pi^0$ using $772 \times 10 ^6 B\bar{B}$ pairs and obtain  
\begin{eqnarray}
\mathcal{S} = -0.92 ~^{+0.31}_{-0.27}~{\rm (stat.)} ~\pm 0.11 ~{\rm (syst.)}, \nonumber \\
\mathcal{A} = 0.28 \pm 0.21~{\rm (stat.)} ~\pm 0.04 ~{\rm (syst.)}. \nonumber  
\end{eqnarray}
The result for $\mathcal{S}$ is consistent with the value measured from decays mediated by a $b \to c \bar{c} s$ transition, $\sin 2 \phi_1 = 0.698 \pm 0.017$ \cite{HFLAV}. The result for $\mathcal{A}$ is consistent with zero, i.e., no direct $CP$ violation, as expected in the SM. This is the first result obtained by the Belle experiment for this mode (and it is the third $CP$-even eigenstate from $b \to s q \bar{q}$ transitions used by Belle for the $\sin 2 \phi_1^{\rm eff}$ measurement after $B^0 \to \eta' K^0_L$ and $B^0 \to \phi K^0_L$).

We thank the KEKB group for the excellent operation of the
accelerator; the KEK cryogenics group for the efficient
operation of the solenoid; and the KEK computer group,
the National Institute of Informatics, and the 
Pacific Northwest National Laboratory (PNNL) Environmental Molecular Sciences Laboratory (EMSL) computing group for valuable computing
and Science Information NETwork 5 (SINET5) network support.  We acknowledge support from
the Ministry of Education, Culture, Sports, Science, and
Technology (MEXT) of Japan, the Japan Society for the 
Promotion of Science (JSPS), and the Tau-Lepton Physics 
Research Center of Nagoya University; 
the Australian Research Council;
Austrian Science Fund under Grant No.~P 26794-N20;
the National Natural Science Foundation of China under Contracts
No.~11435013,  
No.~11475187,  
No.~11521505,  
No.~11575017,  
No.~11675166,  
No.~11705209;  
Key Research Program of Frontier Sciences, Chinese Academy of Sciences (CAS), Grant No.~QYZDJ-SSW-SLH011; 
the  CAS Center for Excellence in Particle Physics (CCEPP); 
Fudan University Grant No.~JIH5913023, No.~IDH5913011/003, 
No.~JIH5913024, No.~IDH5913011/002;                        
the Ministry of Education, Youth and Sports of the Czech
Republic under Contract No.~LTT17020;
the Carl Zeiss Foundation, the Deutsche Forschungsgemeinschaft, the
Excellence Cluster Universe, and the VolkswagenStiftung;
the Department of Science and Technology of India; 
the Istituto Nazionale di Fisica Nucleare of Italy; 
National Research Foundation (NRF) of Korea Grants No.~2014R1A2A2A01005286, No.2015R1A2A2A01003280,
No.~2015H1A2A1033649, No.~2016R1D1A1B01010135, No.~2016K1A3A7A09005 603, No.~2016R1D1A1B02012900; Radiation Science Research Institute, Foreign Large-size Research Facility Application Supporting project and the Global Science Experimental Data Hub Center of the Korea Institute of Science and Technology Information;
the Polish Ministry of Science and Higher Education and 
the National Science Center;
the Ministry of Education and Science of the Russian Federation and
the Russian Foundation for Basic Research;
the Slovenian Research Agency;
Ikerbasque, Basque Foundation for Science, Basque Government (No.~IT956-16) and
Ministry of Economy and Competitiveness (MINECO) (Juan de la Cierva), Spain;
the Swiss National Science Foundation; 
the Ministry of Education and the Ministry of Science and Technology of Taiwan;
and the United States Department of Energy and the National Science Foundation.


\begin{thebibliography}{99}

\bibitem{KM}
M.~Kobayashi and T.~Maskawa, Prog. Theor. Phys. {\bf 49}, 652 (1973).

\bibitem{Sanda}
A.~B.~Carter and A.~I.~Sanda, Phys. Rev. Lett. {\bf 45}, 952 (1980); 
A.~B.~Carter and A.~I.~Sanda, Phys. Rev.  D {\bf 23}, 1567 (1981); 
I.~I.~Bigi and A.~I.~Sanda, Nucl. Phys. {\bf 193}, 85 (1981).

\bibitem{CPVrev}
A general review of the formalism is given in
I. I.~Bigi, V. A.~Khoze, N. G.~Uraltsev, and A. I.~Sanda, ``$CP$ Violation''
page 175, ed. C.~Jarlskog, World Scientific, Singapore (1989). 

\bibitem{plb596_163} T. Gershon and M. Hazumi, Phys. Lett. B {\bf 596} 163 (2004).

\bibitem{beta} Another naming convention, $\beta$ ($= \phi_1$) is also used in the literature.

\bibitem{phi1_belle} I. Adachi {\it et. al.} (Belle Collaboration), Phys. Rev. Lett. {\bf 108}, 171802 (2012).

\bibitem{phi1_babar} B. Aubert, {\it et. al.} (BaBar Collaboration), Phys. Rev. D {\bf 79} 072009 (2009).

\bibitem{phi1_NP} Y. Grossman and M. Woarh, Phys. Lett. B {\bf 395} 241 (1997). 
  
\bibitem{shift_btou} H-. Y. Cheng, arXiv:0702252v1 [hep-ph] (2007).
  
\bibitem{babar2007} B. Aubert, {\it et. al.} (BaBar Collaboration), Phys. Rev. D {\bf 76} 071101 (2007).

\bibitem{Belle}
A.~Abashian {\it et al.} (Belle Collaboration), Nucl. Instr. and Meth. A {\bf 479}, 117 (2002); also see
detector section in J. Brodzicka {\it et al.}, Prog. Theor. Exp. Phys. {\bf 2012}, 04D001 (2012).

\bibitem{svd2} Z. Natkaniec {\it et al.} (Belle SVD2 Group), Nucl. Instr. and Meth. A {\bf 560}, 1(2006).

\bibitem{pi0k0-belle} M. Fujikawa {\it et al.} (Belle Collaboration),  Phys. Rev. D {\bf 81} 011101, (2010). 

\bibitem{ksksks-belle} K. Sumisawa {\it et al.} (Belle Collaboration),  Phys. Rev. Lett. {\bf 95} 061801, (2005). 
  
\bibitem{Evtgen} D. J. Lange {\it et al.}, Nucl. Instr. and Meth.  A {\bf 462}, 152 (2001).

\bibitem{GEANT} R. Brun {\it et al.}, CERN DD/EE/84-1 (1984).
  
\bibitem{neurobayse} M. Feindt and U. Kerzel, The NEUROBAYES neural network package, Nucl. Instr.. Meth. A {\bf 559}, 190 (2006).

\bibitem{nisks} H. Nakano, Ph.D Thesis, Tohoku University (2014) Chapter 4, unpublished, \url{https://tohoku.repo.nii.ac.jp/?action=pages_view_main&active_action=repository_view_main_item_detail&item_id=70563&item_no=1&page_id=33&block_id=38}.

\bibitem{SFW}
 The Fox-Wolfram moments were introduced in
 G.~C.~Fox and S.~Wolfram, Phys. Rev. Lett. {\bf 41}, 1581 (1978).
 The Fisher discriminant used by Belle, based on modified Fox-Wolfram
 moments (SFW), is described in 
 K.~Abe {\it et al.} (Belle Collaboration), Phys. Rev. Lett. {\bf 87},
 101801 (2001) and
 K.~Abe {\it et al.} (Belle Collaboration), Phys. Lett. B {\bf 511}, 151
 (2001). 

\bibitem{KSFW}
S. H. Lee {\it et al.} (Belle Collaboration), Phys. Rev. Lett. {\bf 91},
261801 (2003).

\bibitem{vertexres}
H. Tajima {\it et al.}, Nucl. Instr. and Meth. A {\bf 533} 370 (2004). 

\bibitem{TaggingNIM}
H. Kakuno {\it et al.}, Nucl. Instr. and Meth. A {\bf 533} 516 (2004). 

\bibitem{ARGUS}
H. Albrecht {\it et al.} (ARGUS Collaboration), Phys. Lett. B {\bf 241}, 278 (1990).

\bibitem{PDG}
C. Patrignani {\it et al.} (Particle Data Group), Chin. Phys. C, 40, 100001 (2016) and 2017 update.
  
\bibitem{TSI}
O. Long, M. Baak, R. N. Cahn, and D. Kirkby, Phys. Rev. D {\bf 68} 034010 (2003).

\bibitem{HFLAV} Y. Amhis {\it et al.} (Heavy Flavor Averaging Group), Eur. Phys. J. C 77 895 (2017), arXiv:1612.07233v3 [hep-ex] and online update at https://hflav.web.cern.ch.

\end{thebibliography}
\end{document}